\documentclass[conference]{IEEEtran}
\IEEEoverridecommandlockouts
\usepackage{cite}
\usepackage{amsmath,amssymb,amsfonts}
\usepackage{algorithm,algorithmic}
\usepackage{graphicx}
\usepackage{mysymbol}
\usepackage{textcomp}
\usepackage{xcolor}
\usepackage{url}

\renewcommand{\baselinestretch}{0.94}

\newtheorem{mytheorem}{Theorem}
\newtheorem{myremark}{Remark}
\newtheorem{mydefinition}{Definition}

\def\BibTeX{{\rm B\kern-.05em{\sc i\kern-.025em b}\kern-.08em
    T\kern-.1667em\lower.7ex\hbox{E}\kern-.125emX}}
\begin{document}


\title{Blind Deconvolution of Graph Signals:\\ Robustness to Graph Perturbations
\thanks{This work was supported in part by the Center of Excellence in Data Science, an Empire State Development-designated Center of Excellence.}
}

\author{
\IEEEauthorblockN{
Chang Ye and Gonzalo Mateos
} %
\IEEEauthorblockA{Dept. of ECE, University of Rochester, Rochester, NY, United States} %
}

\maketitle

%

\begin{abstract}
We study blind deconvolution of signals defined on the nodes of an undirected graph. Although observations are bilinear functions of both unknowns, namely the forward convolutional filter coefficients and the graph signal input, a filter invertibility requirement along with input sparsity allow for an efficient linear programming reformulation. Unlike prior art that relied on perfect knowledge of the graph eigenbasis, here we derive stable recovery conditions in the presence of small graph perturbations. We also contribute a provably convergent robust algorithm, which alternates between blind deconvolution of graph signals and eigenbasis denoising in the Stiefel manifold. Reproducible numerical tests showcase the algorithm's robustness under several graph eigenbasis perturbation models.
\end{abstract}



\section{Introduction}
Consider an undirected network graph $\ccalG(\ccalV,\ccalE)$, where $\ccalV$ and $\ccalE$ are the vertex and edge sets, respectively. Neural activities~\cite{weiyu_brain_signals,hu2016localizing,saman2021online}, vehicle trajectories~\cite{deri2016new,gsp2018tutorial,rasoul2021nonstationary}, infectives due to an epidemic~\cite{yang2021covid,segarra2017blind}, temperatures monitored by sensors~\cite{dong2016laplacian,xiao2023sampling}, or productivity of various economic sectors~\cite{antonio2016sampling,hashemi2022greedy}, can all be represented as graph signals $\bbx\in\reals^N$ on $\ccalV$, where $|\ccalV|=N$. Graph Signal Processing (GSP) deals with representations, filters, and algorithms that exploit the graph-induced relational structure of these signals~\cite{gsp2018tutorial,stankovic2019spmag,geert2023spmag}. Building on the graph shift operator (GSO) $\bbS\in\reals^{N\times N}$ that encodes the graph topology, i.e., $S_{ij}\not = 0$ if and only if $(i,j)\in\ccalE$, graph convolutional filters are polynomials $\bbH = \sum_{l = 0}^{L-1}h_l\bbS^l$ that serve as local aggregation operators, with filter coefficients $\bbh=[h_0,\ldots,h_{L-1}]^\top\in \reals^L$~\cite{sandryhaila2013discrete,gama2020spmag,isufi2024gf}. 

In this context, here we revisit the blind deconvolution problem on graphs~\cite{segarra2017blind,chang2018eusipco}. Given $P$ graph signals $\bbY=[\bby_1 \ldots \bby_P]\in\reals^{N\times P}$ that we model as outputs of some degree-$L\leq N$ polynomial graph filter $\bbH$ (where $\bbS$ is perfectly known, for now), 
the goal is to jointly identify the filter taps $\bbh$ and the latent signals $\bbX=[\bbx_1 \ldots \bbx_P]$ driving the observations $\bby_p=\bbH\bbx_p$, up to scaling. 
This bilinear inverse problem is ill-posed, so we assume $\bbX$ is sparse -- say, as when few sources inject a signal that spreads over $\ccalG$~\cite{segarra2017blind}. In this short letter, we study the robustness of existing and new blind deconvolution algorithms in the presence of graph perturbations.\vspace{2pt}
%

\noindent \textbf{Related work.} Our starting point is a convex relaxation put forth in~\cite{chang2018eusipco}, feasible under a mild invertibility assumption on $\bbH$ (Section \ref{sec:Prelim}) and inspired by the (graph-agnostic) blind deconvolution method in~\cite{wang2016blind}. The algorithm in~\cite{chang2018eusipco} outperforms its matrix-lifting precursor~\cite{segarra2017blind}, and comes with exact recovery guarantees as well as noise stability bounds under the Bernoulli-Gaussian model for $\bbX$~\cite{chang2024exact}. However,~\cite{segarra2017blind,chang2018eusipco,chang2024exact} rely on perfect knowledge of the GSO (or its eigenbasis), and performance under pragmatic graph perturbations is yet to be explored. Resilience to imperfect topology information has received attention in other GSP contexts; noteworthy contributions include small perturbation analyses~\cite{ceci2018small,ceci2020graph}, stability properties of graph neural networks (GNNs)~\cite{gama2020stability}, and topology denoising under signal and connectivity errors~\cite{ceci2020tls,samuel2023robust}. 
Most germane to our work,~\cite{victor2024icassp} proposed a node-domain, non-convex blind deconvolution approach from multiple sparse signals on graphs with edge perturbations. 
Here, instead, we operate in the graph frequency domain and prove the method in~\cite{chang2018eusipco} (which inspired~\cite{victor2024icassp} as well) is stable; see Section \ref{sec:stability}.\vspace{2pt}

\noindent \textbf{Contributions.} We derive a new stability result that ensures the filter estimation error remains manageable even when the eigenvectors of $\bbS$ (i.e., the graph Fourier basis) exhibit small perturbations. Modeling an imperfect graph eigenbasis is well motivated when e.g., $\ccalG$ is unknown, but eigenvectors can still be estimated from $P$ stationary signals in $\bbY$; see~\cite{segarra2017network} and Remark \ref{rem:covariance}.  
Our second main contribution is a robust blind deconvolution formulation and provably convergent optimization algorithm in the Stiefel manifold~\cite{peng2023block}, to jointly ``denoise'' the given perturbed graph eigenbasis (Section \ref{sec:algorithm}). Numerical tests showcase the method's enhanced robustness relative to the blind deconvolution algorithms in~\cite{chang2018eusipco,victor2024icassp}. In the interest of reproducible research, we share the code used to generate all the figures in Section \ref{sec:numerical}. Concluding remarks are in Section \ref{sec:conclusions}. Due to page constraints, the proof of our main result is deferred to the accompanying Supplementary Material.


\section{Preliminaries and Problem Statement}\label{sec:Prelim}

We begin by reviewing a convex relaxation approach to blind deconvolution on graphs, and state the exact recovery conditions derived in~\cite{chang2024exact}. We then formulate and motivate the new problem arising when graph knowledge is imperfect.\vspace{2pt}
%


\noindent\textbf{Convex relaxation for invertible graph filters.} Similar to~\cite{chang2018eusipco,chang2024exact,victor2024icassp}, we henceforth assume the forward convolutional filter $\bbH$ is invertible. In such case, one can show that the inverse operator $\bbG := \bbH^{-1}$ is also a graph filter on $\ccalG$, which can be uniquely represented as a polynomial in $\bbS$ of degree at most $N-1$~\cite[Theorem 4]{sandryhaila2013discrete}. To be more specific, let $\bbg \in \reals^{N}$ be the vector of inverse-filter coefficients, i.e., $\bbG= \sum_{l=0}^{N-1} g_l \bbS^l$. Since $\bbS$ is symmetric, it is diagonalizable as $\bbS=\bbV\bbLambda\bbV^\top$, with $\bbLambda=\textrm{diag}(\lambda_1,\ldots,\lambda_N)$ collecting the eigenvalues and the eigenvectors $\bbV$ are akin to Fourier modes~\cite{gsp2018tutorial}. Accordingly, we equivalently rewrite the forward model $\bbY=\bbH\bbX$ as $\bbX  = \bbG \bbY  = \bbV \text{diag}(\tbg) \bbV^\top \bbY$, where $\tilde{\bbg} := \bbPsi_N \bbg \in \reals^N$ is the inverse filter's frequency response and $\bbPsi_N \in \reals^{N \times N}$ is Vandermonde with $\Psi_{ij}:=\lambda_i^{j-1}$~\cite{isufi2024gf}. 

Leveraging the invertibility of $\bbH$ and exploiting sparsity via an $\ell_1$-norm criterion,~\cite{chang2018eusipco} recasts the blind deconvolution task as a \emph{convex and linear} inverse problem 
\begin{equation}\label{pb:exact}
\hat{\tbg}  = \text{arg}\min\limits_{\tbg} \| \bbV\diag{(\tbg)}\bbV^\top\bbY\|_{1,1}
,\quad \text{s. to }\:
\mathbf{1}_N^\top\tbg = N,
\end{equation}
where $\|\bbX\|_{1,1} = \sum_{ij}|X_{ij}|$. The constraint in \eqref{pb:exact} excludes $\hat{\tbg}=\mathbf{0}_N$ and fixes the (arbitrary) scale of the solution; see~\cite{chang2024exact}.
Theoretical analysis of the convex estimator \eqref{pb:exact} is possible with the following tractable model for random sparse matrices.
\begin{mydefinition}[Bernoulli-Gaussian model]\label{def:BG} We say a random matrix $\bbX\in\reals^{N\times P}$ adheres to the Bernoulli-Gaussian model with parameter $\theta\in(0,1)$, if $X_{ip}  = \Omega_{ip}\gamma_{ip}/\sqrt\theta$, where $\Omega_{ip}  \sim\text{Bernoulli}(\theta)$ and $\gamma_{ip} \sim \text{Normal}(0,1)$ are i.i.d. for all $i,p$.
\end{mydefinition}

Let $\bbP_1^\perp := \bbI_N-\frac{\mathbf{1}_N\mathbf{1}_N^\top}{N}$ be the projection onto $\textrm{span}^{\perp}(\mathbf{1}_N)$ and define matrix $\tbU := (\bbV \circ \bbV)\bbP_1^\perp\in\reals^{N\times N}$, where $\circ$ denotes the Hadamard product. One can show the spectral radius $\sigma_{\max}(\tbU)\leq 1$. The ensuing theorem offers a sufficient condition under which \eqref{pb:exact} succeeds with high probability.
\begin{mytheorem}[Exact recovery~\cite{chang2024exact}]\label{theorem_exact}
\normalfont Consider graph signal observations $\bbY = \bbV\diag(\tbh_0)\bbV^\top\bbX_0\in\reals^{N\times P}$, where $\bbX_0$ adheres to the Bernoulli-Gaussian model with $\theta\in \left(0,0.324\right]$ and $\tbh_0 \circ \tbg_0=\mathbf{1}_{N}$. 
Let $P \geq C'\sigma_m^{-2}\log{\frac{4}{\delta}}$, where $\sigma_m = \min(\sigma_1,\sigma_2,\sigma_3,\sigma_4)$ and $\sigma_1\in \left(0,\frac{\sqrt{\pi}\theta^{3/2}}{2}\right]$, $\sigma_2 \in \left(0,\frac{\sqrt{\pi}\theta}{2}\right]$, $\sigma_3>0$, $\sigma_4\in(0,1)$,   $\delta\in(0,1)$ are parameters, while $C'$ is a constant that does not depend on $P$, {$\sigma_m$}, or $\delta$. Then $\hat\tbg = \tilde{\bbg}_0$ is the unique solution to \eqref{pb:exact} with probability at least $1-\delta$, if
\begin{equation} \label{eq:reco_gaur}
\left\|\mathbf{P}^\perp_1\Tilde{\bbg}_0\right\|_2  \leq a_0,
\end{equation}
where $a_0 := \frac{\sqrt{1 - \sigma_{\max}^2(\Tilde{\bbU})}\left[(1-\sigma_1)-2\theta(1+\sigma_2)\right](1-\sigma_4)}{(1+\sigma_3)\sqrt\theta}$. 
\end{mytheorem}
From \eqref{eq:reco_gaur}, one can interpret $\left\|\mathbf{P}^\perp_1\Tilde{\bbg}_0\right\|_2$ as a measure of the filter's (and the problem's) ill-conditioning. Moreover, increasing $\theta$ within its feasible range will decrease $a_0$, thus challenging recovery. Please refer to~\cite{chang2024exact} for an expanded discussion. \vspace{2pt} 

\noindent\textbf{Problem statement.} Going beyond prior algorithmic~\cite{chang2018eusipco,victor2024icassp} and theoretical studies~\cite{chang2024exact,segarra2017blind} of blind deconvolution on graphs, in this letter we investigate the robustness of \eqref{pb:exact} when the knowledge of $\ccalG$ is imperfect. There have been several graph perturbation models considered in the GSP literature\cite{ceci2018small,ceci2020graph,ceci2020tls,gama2020stability,samuel2023robust,victor2024icassp}, but given the spectral domain approach \eqref{pb:exact} we directly focus on perturbations $\bbDelta=\bbV-\bbV_p$ to the graph's eigenbasis $\bbV$. Given observations $\bbY = \bbV\diag(\tbh)\bbV^\top\bbX$ and \emph{perturbed} GSO eigenvectors $\bbV_p$ ($\bbS$ is no longer perfectly known), we ask what is the performance of [cf. \eqref{pb:exact}]
%
%
\begin{equation} \label{pb:perturbed}
\hat{\tbg}_p  = \text{arg}\min\limits_{\tbg} \| \bbV_p\diag{(\tbg)}\bbV_p^\top\bbY\|_{1,1}
,\quad \text{s. to }\:
\mathbf{1}_N^\top\tbg = N.
\end{equation}
\begin{myremark}[Motivating graph eigenbasis perturbations]\label{rem:covariance}
Even when the graph topology $\bbS$ is completely unknown, under the Bernoulli-Gaussian model (Definition \ref{def:BG}) one can still (imperfectly) estimate the graph eigenvectors from the observations $\bbY=\bbH\bbX$. Indeed, since $\bbC_x:=\E{\bbx_p\bbx_p^\top}=\bbI_N$, the ensemble covariance matrix of the observations is 
\begin{equation*}
\bbC_y:=\E{\bby_p\bby_p^\top} = \bbH\E{\bbx_p\bbx_p^\top} \bbH^\top = \bbV\diag^2{(\tbh)}\bbV^\top.
\end{equation*}
The conclusion is that the GSO eigenvectors coincide with those of $\bbC_y$, and one can form an estimate $\bbV_p$ by diagonalizing the sample covariance matrix $\hbC_y=\frac{\bbY\bbY^\top}{P-1}$; see also~\cite{segarra2017network}.
%
%
\end{myremark}


\section{Stable Recovery Under Graph Perturbations}\label{sec:stability}


Our main stable recovery result asserts that the estimation error on the inverse filter’s frequency response can be kept at a manageable level, even when the given GSO eigenbasis is subject to small errors $\bbDelta=\bbV-\bbV_p$. Recalling the statement and conditions of Theorem \ref{theorem_exact}, let $\ccalS:=\textrm{supp}(\bbX_0)=\textrm{supp}(\bbOmega)$. Defining $\bbE:= (\bbV -\bbDelta)[\bbDelta^\top - \diag(\tbg_0)\bbDelta^\top\bbH_0]\bbX_0\in\reals^{N\times P}$, let $\bbE^{(\ccalS^C)}:=\bbE\circ (\mathbf{1}_{N\times N}-\bbOmega)$ be the restriction to the entries in the complement $\ccalS^C$ of the support of $\bbX_0$. We can thus establish the following error bound for the solution of \eqref{pb:perturbed}.  
%
%
\begin{mytheorem}[Stable recovery to graph perturbations] \label{theorem_1}
\normalfont
Consider the same setting and conditions in Theorem \ref{theorem_exact}, including \eqref{eq:reco_gaur} for $a_0\geq 0$ that does not depend on $\bbDelta$. Then the estimation error associated to the solution $\hat\tbg_p$ of \eqref{pb:perturbed}
is bounded by
\begin{equation} \label{eq:thm}
  \left\|\hat\tbg_p - \tbg_0\right\|_2 \leq\frac{2\sigma_\textrm{max}\left(\diag(\tbg_0) - \frac{\tbg_0\tbg_0^\top}{N}\right)\|\bbE^{(\ccalS^C)}\|_{1,1}}{PQ - a_0  \|\bbE^{(\ccalS^C)}\|_{1,1}  - \|[\bbE^{(\ccalS^C)}]^\top\bbV\odot\bbV\|_{1\rightarrow 2}}, 
\end{equation}
where $Q:=C_1\sqrt{\theta}(\sqrt{a_0^2-(1-\sigma)^2\|\bbP_1^\perp \Tilde{\bbg}_0\|_2^2}-\sigma \|\bbP_1^\perp \Tilde{\bbg}_0\|_2)$ for some $\sigma\in[0,1]$ and $C_1>0$. The operator norm $\|\cdot\|_{1\rightarrow 2}$ denotes the maximum $\ell_2$-norm for the columns of its matrix argument, and $\odot$ stands for the Kathri-Rao product. 
\end{mytheorem}
\begin{myproof}
Included in the Supplementary Material.
\end{myproof}

Notice first that $Q\geq 0$, because $\|\bbP_1^\perp \Tilde{\bbg}_0\|_2\leq a_0$ as per \eqref{eq:reco_gaur}. Importantly, the denominator in the right-hand side of \eqref{eq:thm} should be non-negative to obtain a feasible upper bound. This effectively imposes a constraint on the magnitude of the error component $\bbE^{(\ccalS^C)}$, which should satisfy 
\begin{equation}\label{eq:thm_2}
    \|\bbE^{(\ccalS^C)}\|_F \leq PQ/M_1,
\end{equation}
where $M_1:=a_0\|\bar{\bbE}^{(\ccalS^C)}\|_{1,1} + \|[\bar{\bbE}^{(\ccalS^C)}]^\top\bbV\odot\bbV\|_{1\rightarrow 2}$ and $\bar{\bbE}^{(\ccalS^C)} := \bbE^{(\ccalS^C)}/\|\bbE^{(\ccalS^C)}\|_F$. A sufficient condition for \eqref{eq:thm_2} to hold is that $\|\bbE\|_F\leq PQ/M_1$, since $\|\bbE^{(\ccalS^C)}\|_F\leq \|\bbE\|_F$. Because $\|\bbE\|_F$ is proportional to the magnitude of $\bbDelta$, namely
\begin{equation*}
    \|\bbE\|_F = \|[\bbDelta^\top - \diag(\tbg_0)\bbDelta^\top\bbH_0]\bbX_0\|_F = M_2\|\bbDelta\|_F,
\end{equation*}
where $M_2:=\|[\bar\bbDelta^\top - \diag(\tbg_0)\bar\bbDelta^\top\bbH_0]\bbX_0\|_F$ and $\bar\bbDelta := \bbDelta/\|\bbDelta\|_F$, then we can upper bound $\|\bbDelta\|_F$ as
\begin{equation} \label{eq:thm_3}
    \|\bbDelta\|_F \leq PQ/(M_1 M_2).
\end{equation}
The right-hand side of \eqref{eq:thm_3} provides an upper bound to the magnitude of the eigenbasis perturbation that is tolerable. Again, in the favorable setting where $\|\bbP_1^\perp \Tilde{\bbg}_0\|_2$ is small, e.g., if $\tbg_0$ is closer to the all-ones vector $\mathbf{1}_N$, we will have a larger upper bound in \eqref{eq:thm_3} because $Q$ increases and $M_2$ decreases. 
%


\section{Robust Blind Deconvolution Algorithm}\label{sec:algorithm}

Theorem \ref{theorem_1} shows the convex relaxation \eqref{pb:exact} is robust to small errors $\bbDelta$ in the GSO eigenbasis, complementing the noise stability results in~\cite{chang2024exact}. Here we develop a new robust blind deconvolution formulation and associated algorithm to jointly estimate $\tbg$ and denoise $\bbV_p$. In Section \ref{sec:numerical} we empirically show it outperforms \eqref{pb:perturbed}, especially for larger perturbations. 

Our idea is to adopt the Huber loss~\cite{Huber_Loss} (with small $\epsilon\geq 0$)
\begin{equation}
     h_{\epsilon}(x) :=\left\{
    \begin{aligned}
        \frac{x^2}{2\epsilon},\:\: |x| < \epsilon\\
        |x| - \frac{\epsilon}{2},\:\: |x| \geq \epsilon
    \end{aligned}
    \right.
\end{equation} 
in order to construct a smooth surrogate $f(\tbg,\bbV):=\sum_{i,j}h_{\epsilon}(\bbe_i^\top\bbV\diag(\tbg)\bbV^\top\bby_j)$ to the cost function in \eqref{pb:exact}, where $\bbe_i$ is the $i$-th coordinate vector. Note that $f(\tbg,\bbV)$ is differentiable with a Lipschitz gradient and $f(\tbg,\bbV) \leq \Vert\bbV\diag(\tbg)\bbV^\top\bbY\rVert_{1,1}\leq f(\tbg,\bbV) + \frac{\epsilon}{2}$. Accordingly, we propose to solve the following smooth, manifold-constrained problem
%
%
%
%
%
%
\begin{equation} \label{pb:smooth}
\min_{\tbg,\bbV\in \ccalM_{St}} \Big\{ \underbrace{f(\tbg,\bbV)+ \frac{\rho}{2}\|\bbV-\bbV_p\|_F^2}_{:=F(\tbg,\bbV)}\Big\},\:\text{ s. to } \mathbf{1}_N^\top\tbg = N,
\end{equation}
where $\ccalM_{St} = \{\bbV\in\reals^{N\times N}\:|\:\bbV^\top\bbV=\bbI_N\}$ is the Stiefel manifold of orthogonal matrices. Importantly, we \emph{jointly} optimize over the inverse filter frequency response and the denoised graph eigenvectors, using a criterion $F(\tbg,\bbV)$ that combines data consistency and encourages similarity to the given $\bbV_p$.
%


%
%

\begin{figure}[t] 
		\centering
	\includegraphics[width=0.9\linewidth]{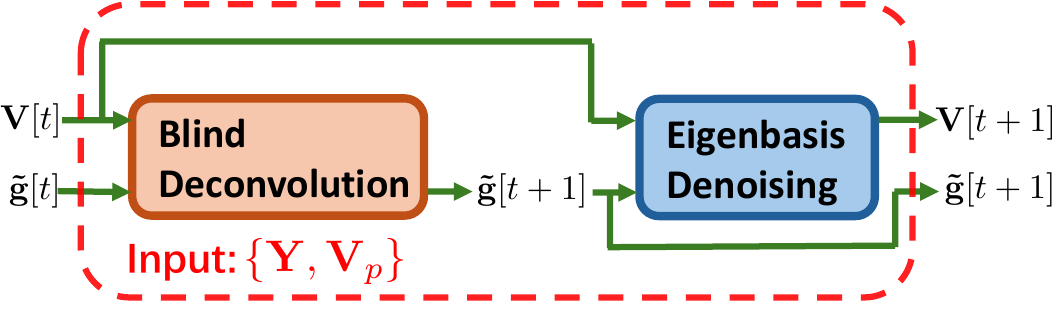}
 \caption{Schematic diagram of an iteration of the robust blind deconvolution of graph signals (RBDoGS) algorithm.} \label{f:illustrate}
\end{figure}

Solving \eqref{pb:smooth} is challenging since the problem is non-convex. Yet its structure lends itself naturally to an iterative block-coordinate descent approach in the Stiefel manifold~\cite{peng2023block}. At iteration $t=0,1,2,\ldots$ we blend: (i) block exact minimization of \eqref{pb:smooth} w.r.t. $\tbg$ for fixed $\bbV[t]$ to obtain $\tbg[t+1]$; and (ii) then perform one step of block Riemannian gradient descent (BRGD) w.r.t. $\bbV$ to update $\bbV[t+1]$. The  update steps (i)-(ii) are detailed below and also schematically illustrated in Fig. \ref{f:illustrate}.\vspace{2pt} 

\noindent\textbf{Blind deconvolution.} At step (i) of iteration $t$, we compute a minimizer $\tbg[t+1]$ of $F(\tbg,\bbV[t])$ by solving the smoothed convex blind deconvolution subproblem,
\begin{equation} \label{pb:source_loc_smooth}
\tbg[t+1] = \argmin_{\tbg} f(\tbg,\bbV[t]),\:
 \text{ s. to }\: \mathbf{1}_N^\top\tbg = N,
\end{equation}
where $\bbV[t]$ is the eigenbasis estimate from iteration $t-1$. In practice, we can also obtain $\tbg[t+1]$ by solving \eqref{pb:perturbed} with $\bbV_p = \bbV[t]$, since  \eqref{pb:perturbed} and \eqref{pb:source_loc_smooth} are equivalent when $\epsilon\rightarrow 0$.\vspace{2pt} 

\noindent \textbf{Eigenbasis denoising.} At step (ii), computational considerations motivate applying a single iteration of BRGD to update 
\begin{equation} \label{pb:eigen_denois_smooth}
\bbV[t+1] = \textrm{Retr}_{\bbV[t]}(-\beta_t\tilde{\nabla}_\bbV F(\tbg[t+1],\bbV[t])), \\
\end{equation}
where $\textrm{Retr}_{\bbV[t]}(\cdot)$ is the retraction mapping at $\bbV[t]$, $\tilde{\nabla}_\bbV F(\cdot,\cdot)$ is the Riemannian gradient w.r.t. $\bbV$, and $\beta_t$ is a step size. Update \eqref{pb:eigen_denois_smooth} can be computed via the Cayley transform $\bbV[t+1]  = \ccalC(\bbM[t])\bbV[t]$; see e.g.,~\cite[Example 3.6.2]{absil2008optimization}, where
\begin{equation} \label{eq:cayley_trans}
    \begin{aligned}
    \bbG[t] &=\nabla_\bbV F(\tbg[t+1],\bbV[t]),\\
       \bbM[t] &=\bbG[t] \bbV^\top[t] - \bbV[t]\bbG^\top[t],\\
       \ccalC(\bbM[t]) & = \Big(\mathbf{I}_N + \frac{\beta_t}{2}\bbM[t]\Big)^{-1}\Big(\mathbf{I}_N - \frac{\beta_t}{2}\bbM[t]\Big).
    \end{aligned}
\end{equation}
The procedure in \eqref{eq:cayley_trans} connects the Euclidean gradient $\bbG[t]$ of $F(\cdot,\cdot)$ and its BRGD direction that ensures $\bbV[t+1]\in\ccalM_{St}$. We determine the step size $\beta_t$ via line search~\cite{boumal2023intromanifolds}. The pseudo-code of the novel algorithm for robust blind deconvolution of graph signals (RBDoGS) is tabulated under Algorithm \ref{alg:smooth}.

The computational complexity of RBDoGS is $\ccalO(N^3)$ per iteration. Because \eqref{pb:source_loc_smooth} has a unique minimizer and $F(\cdot,\cdot)$ is block-$i$ Lipschitz smooth \cite[Definition 4]{peng2023block} w.r.t. $\bbV$, then convergence of Algorithm \ref{alg:smooth} follows from~\cite[Theorem 4]{peng2023block}. 

\begin{algorithm}[t]
	\caption{Robust blind deconvolution of graph signals.}
	\label{alg:smooth}
	\begin{algorithmic}[1]
		\STATE 	\textbf{Input: } $\bbY$, $\bbV_p$, $\delta > 0$ and  $\epsilon > 0$.
		\STATE \textbf{Initialize} $t=0$, $\bbV[0]=\bbV_p$,  $\tbg[0] = \mathbf{1}_{N}$.
		\REPEAT
		\STATE Update $\tbg[t]$ by solving \eqref{pb:source_loc_smooth}.\\
        \STATE Update $\bbV[t]$ via BRGD in \eqref{pb:eigen_denois_smooth}; e.g. using \eqref{eq:cayley_trans}.\\
		\STATE $t \gets t+1$.\\
		\UNTIL $\|\tbg[t]- \tbg[t-1]\|_{2} \le \delta$ and $\|\bbV[t] - \bbV[t-1]\|_{2} \le \delta$.\\
		\RETURN $\hat{\tilde{\bbg}}:= \tilde{\bbg}^{(t)}$ and $\hat{\bbX}:=\bbV[t]\diag(\tbg[t])\bbV[t]^\top\bbY$.
	\end{algorithmic}
\end{algorithm}
%

\section{Numerical Experiments}\label{sec:numerical}
\begin{figure*}[ht] \label{fig:main_result}
	\begin{minipage}[b]{0.24\textwidth}
		\centering
		\includegraphics[width=1.1\linewidth]{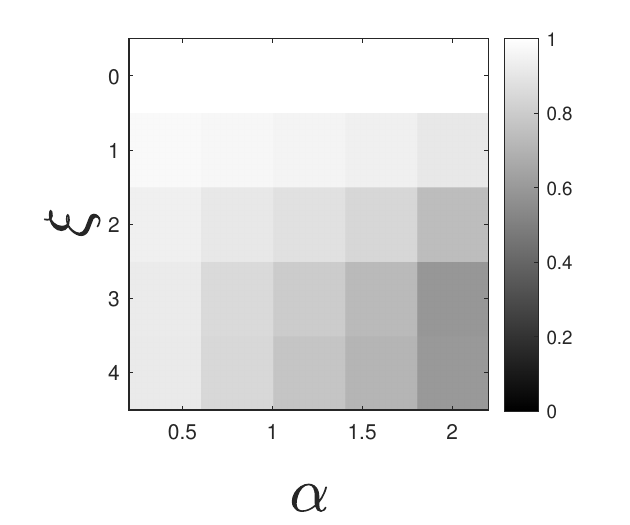}
		\centerline{(a)}\medskip
	\end{minipage}
	\hfill
	\begin{minipage}[b]{0.24\textwidth} 
		\centering
		\includegraphics[width=1.1\linewidth]{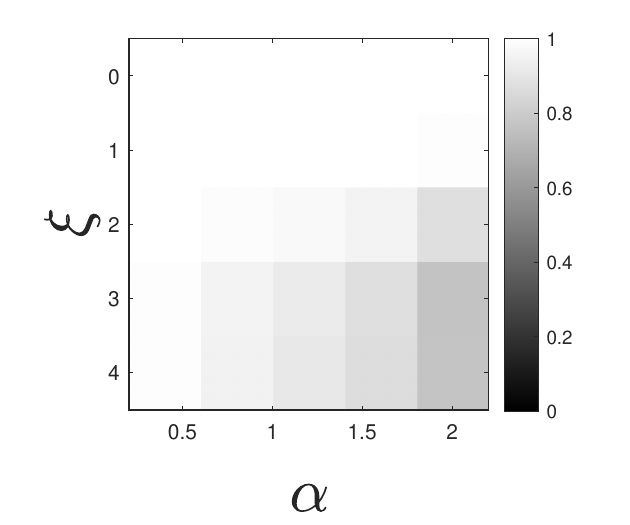}
		\centerline{(b)}\medskip
	\end{minipage}
        \hfill
	\begin{minipage}[b]{0.24\textwidth}
		\centering
		\includegraphics[width=1.1\linewidth]{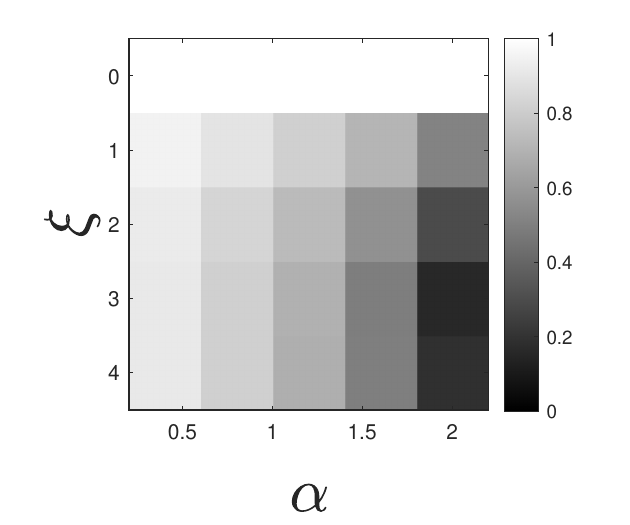}
		\centerline{(c)}\medskip
	\end{minipage}
	\hfill
	\begin{minipage}[b]{0.24\textwidth} 
		\centering
		\includegraphics[width=1.1\linewidth]{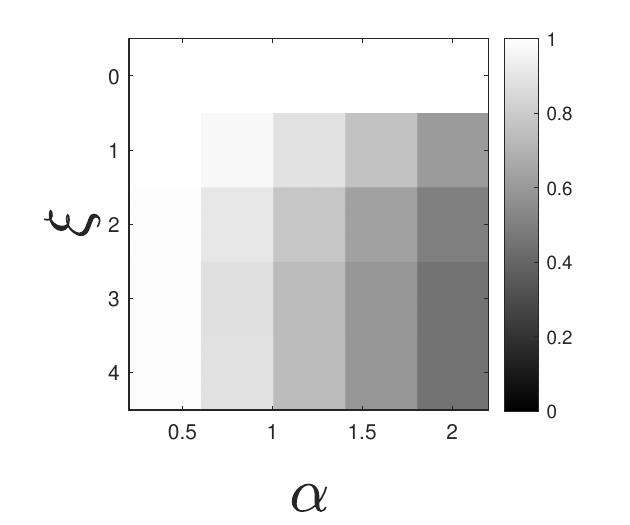}
		\centerline{(d)}\medskip
	\end{minipage}
 \caption{Improved recovery performance of the RBDoGS algorithm on Erdös-Renyi random graphs with $N = 20,\:P=60,\:p_{\ccalE}=0.4$, as a function of $\alpha=\|\bbP^\perp_1\tbg_0\|_2$ and $\xi=\|\bbDelta\|_F$. (a) and (b) show $1-\textrm{RE}_g$ and $\textrm{ACC}_X$ for Algorithm \ref{alg:smooth} (the whiter the better). (c) and (d) show $1-\textrm{RE}_g$ and $\textrm{ACC}_X$ for the BDoG baseline that solves \eqref{pb:perturbed} with an imperfect $\bbV_p$~\cite{chang2018eusipco}. Results depict averages over 100 independent realizations.} \label{f:figure_main}
\end{figure*}
\begin{figure*}[ht] \label{fig:corr_network}

	\begin{minipage}[b]{0.24\textwidth}
		\centering
		\includegraphics[width=1\linewidth]{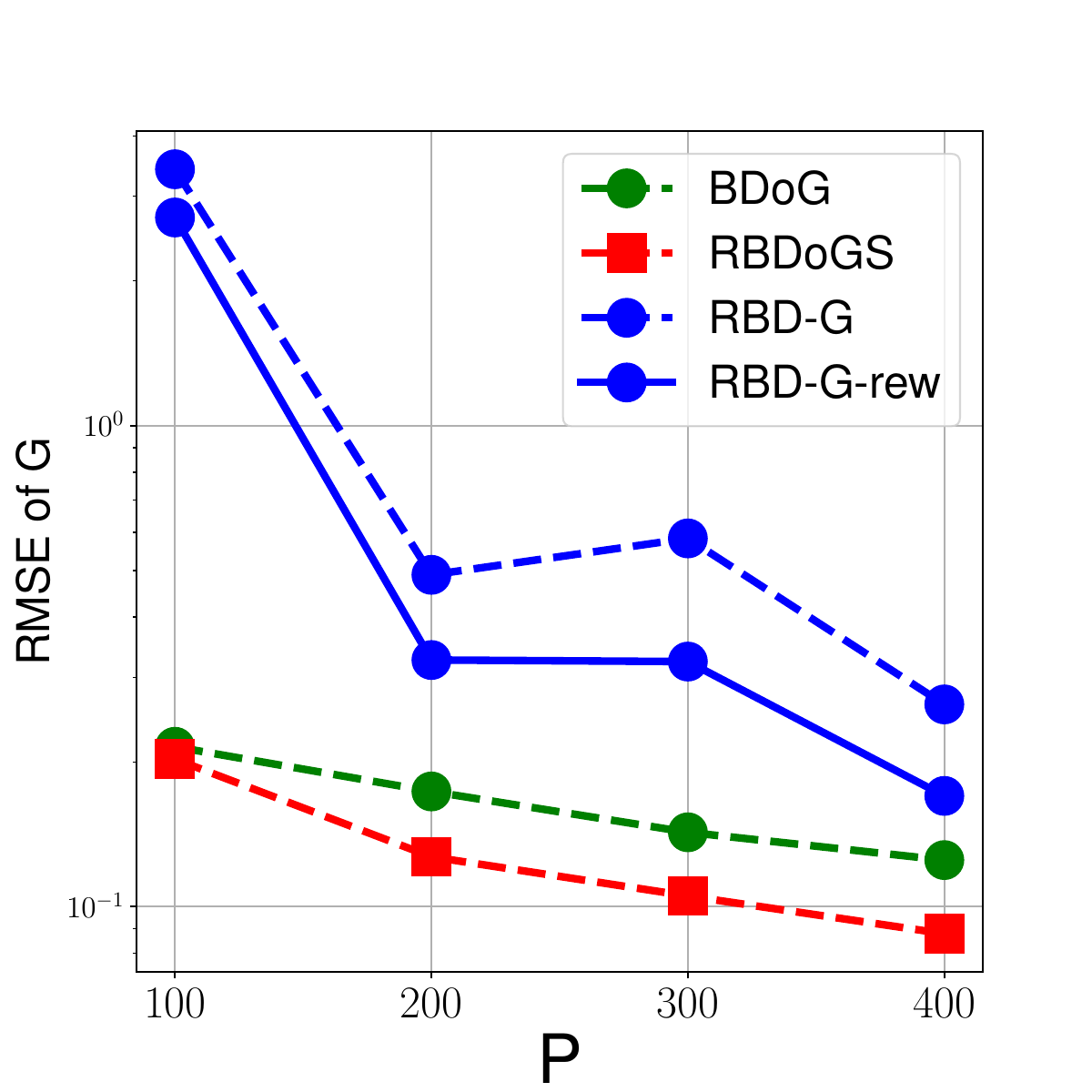}
		\centerline{(a)}\medskip
	\end{minipage}
	\hfill
	\begin{minipage}[b]{0.24\textwidth} 
		\centering
		\includegraphics[width=1\linewidth]{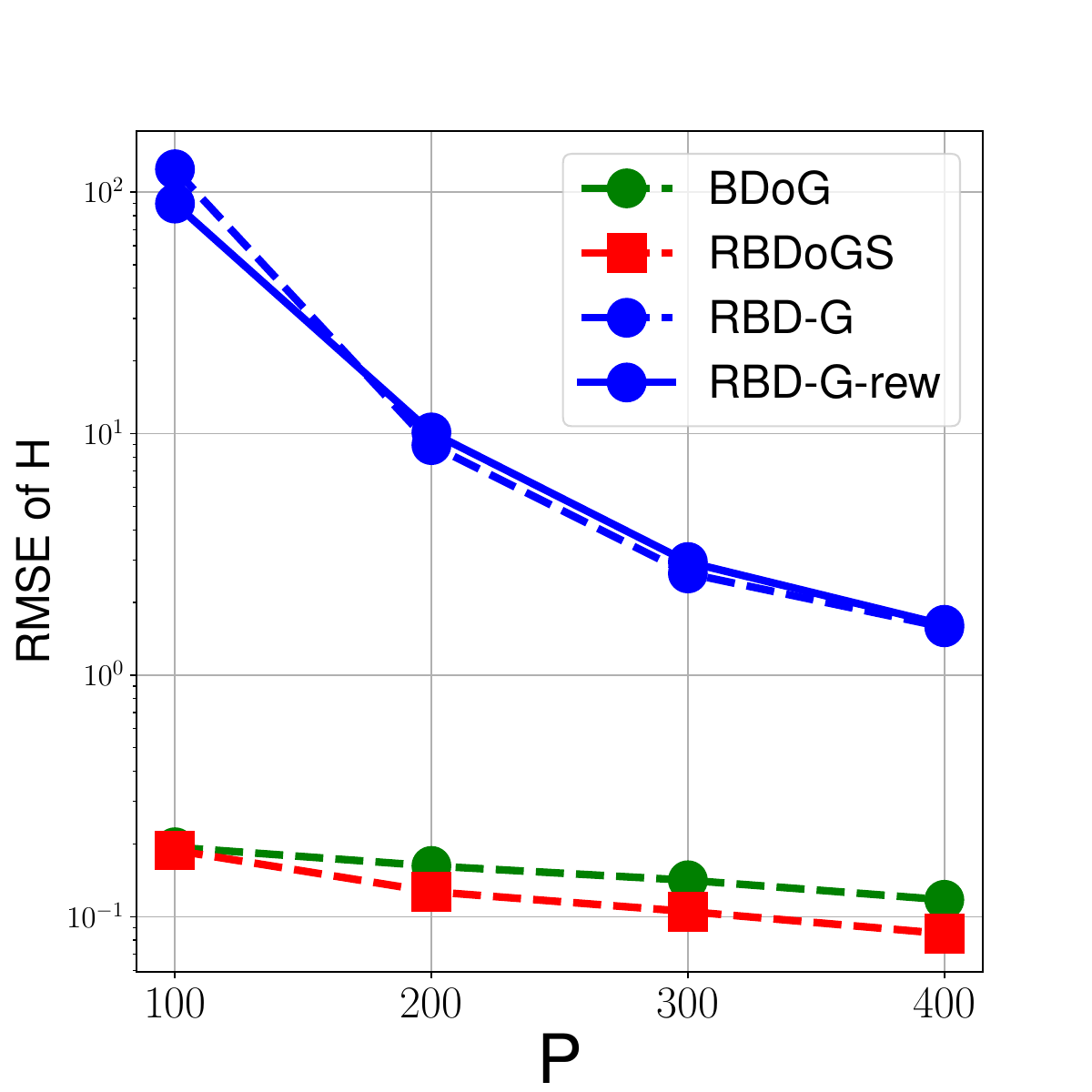}
		\centerline{(b)}\medskip
	\end{minipage}
        \hfill
	\begin{minipage}[b]{0.24\textwidth}
		\centering
		\includegraphics[width=1\linewidth]{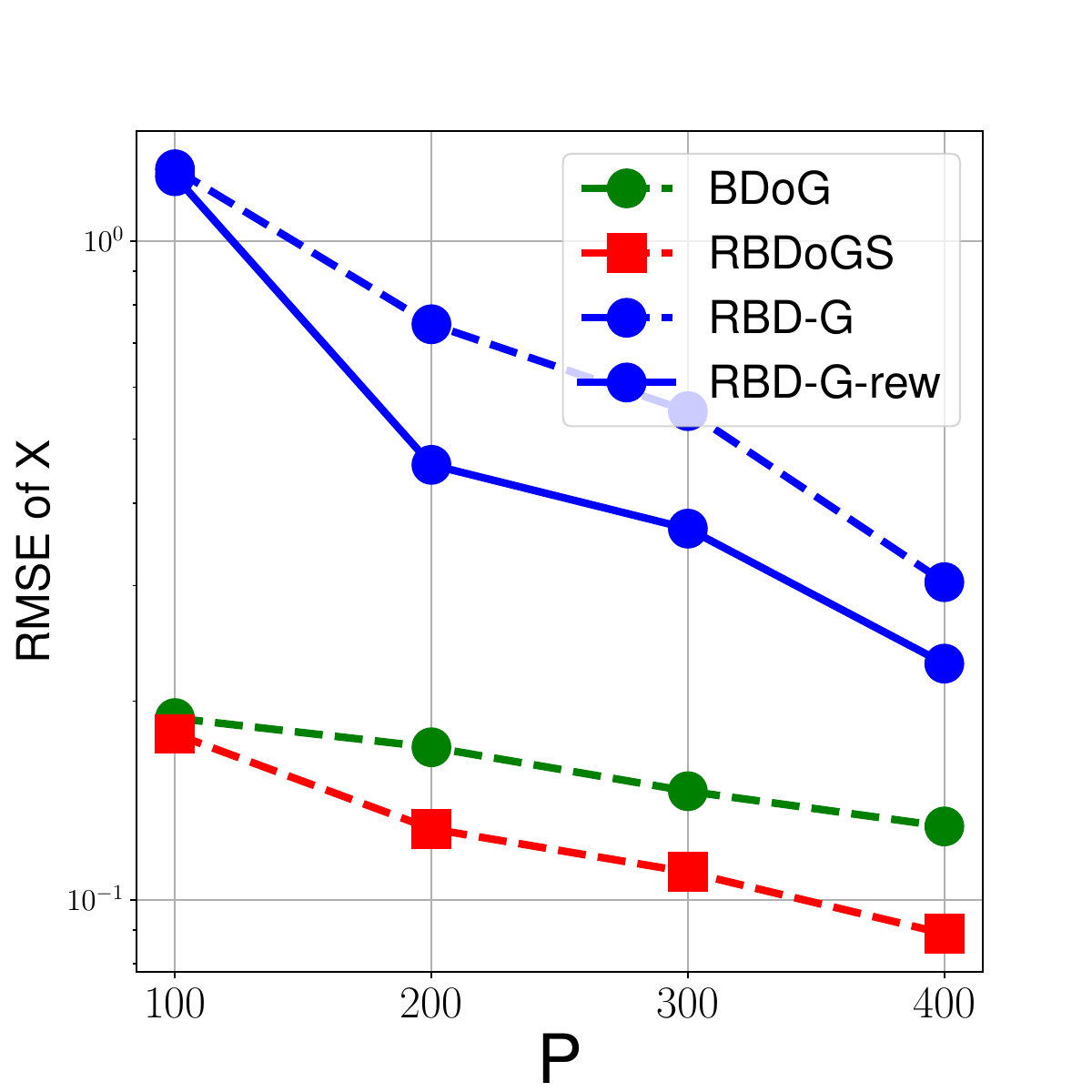}
		\centerline{(c)}\medskip
	\end{minipage}
	\hfill
	\begin{minipage}[b]{0.24\textwidth} 
		\centering
		\includegraphics[width=1\linewidth]{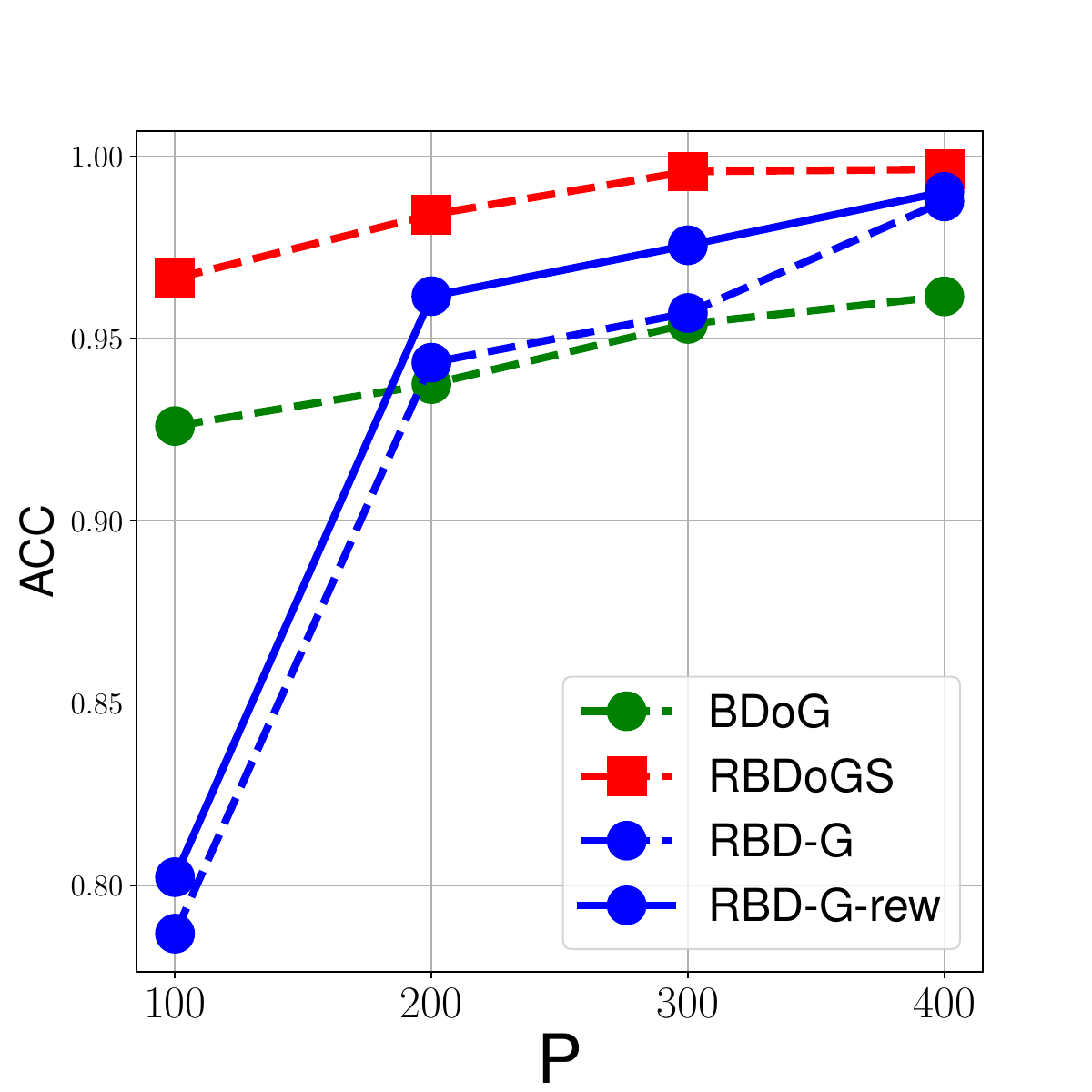}
		\centerline{(d)}\medskip
	\end{minipage}   
    
 \caption{Improved recovery performance of the RBDoGS algorithm on Erdös-Renyi random graphs with $N = 20,\:p_{\ccalE}=0.4$, as a function of the sample size $P$. (a) - (c) show the RMSE of $\bbG$, $\bbH$, $\bbX$, respectively; and (d) shows $\textrm{ACC}_X$ for four competing approaches, including the BDoG baseline that solves \eqref{pb:perturbed} with an imperfect $\bbV_p$~\cite{chang2018eusipco}, the proposed RBDoGS, the RBD-G algorithm and its variant RBD-G-rew in~\cite{victor2024icassp}. Median results over 20 independent realizations.}\label{f:corr_network}
\end{figure*}

For all the experiments in this section we consider Erdös-Renyi random graphs with $N = 20$ nodes and edge formation probability $p_{\ccalE} = 0.4$. We select the degree-normalized adjacency matrix as the GSO, i.e. $\bbS = \bbD^{-1/2}\bbA\bbD^{-1/2}$, where $\bbD=\textrm{diag}(\bbA\mathbf{1}_N)$. Implementation details are in the public code\footnote{\scriptsize{The code is available from: \texttt{\url{https://hajim.rochester.edu/ece/sites/gmateos/code/RBDoGS.zip}}}.}, which can be run to generate all plots in Figs. \ref{f:figure_main} and \ref{f:corr_network}.\vspace{2pt}

\noindent\textbf{Test case 1: Controllable $\|\bbP^\perp_1\tbg_0\|_2$ and $\|\bbDelta\|_F$.} We draw inputs $\bbX_0\in\reals^{N\times P}$ from the Bernoulli-Gaussian model, with $P = 60$ and $\theta = 0.15$. The graph filter $\bbH_0=\bbV\diag(\tbh_0)\bbV^\top$ is generated such that $\tbg_0 = \mathbf{1}_N + (\alpha\bbP_1^\perp\bbb)/\|\bbP_1^\perp\bbb\|_2$, where $\bbb\sim\textrm{Normal}(\mathbf{0}_N,\bbI_N)$. This way we can control $\|\bbP^\perp_1\tbg_0\|_2 = \alpha$, and larger $\alpha$ leads to a harder blind deconvolution problem [cf. \eqref{eq:reco_gaur}]. The observations are generated as $\bbY = \bbH_0\bbX_0$.

In order to simulate graph perturbations with controllable $\|\bbDelta\|_F$, we apply the parameterized Cayley transformation $\ccalC(\bbW,\xi):= (\bbI_N+\xi\bbW)^{-1}(\bbI_N - \xi\bbW)$ to $\bbV$, with parameter $\xi\in[0,\infty)$ and a skew-symmetric matrix $\bbW$ with unit Frobenius norm.  We generate $\bbV_p=\ccalC(\bbW,\xi)\bbV$. The eigenvalues of $\bbW$ are all imaginary, i.e., $\{i\mu_k\}_{k=1}^N$. Then one can derive
\begin{equation} \label{eq:pertu_char}
    \|\bbDelta\|_F^2 = \|[\bbI_N - \ccalC(\bbW,\xi)]\bbV\|_F^2  = \sum_{k=1}^N\frac{4\mu_k^2}{\frac{1}{\xi^2}+\mu_k^2},
\end{equation}
so $\|\bbDelta\|_F^2$ varies from 0 to $4N$ when $\xi$ varies from 0 to $\infty$.
%

We run Algorithm \ref{alg:smooth} with inputs $\{\bbY,\bbV_p\}$, for different values of $\alpha$ and $\xi$. To assess the robustness of RBDoGS, we use the convex blind deconvolution on graphs (BDoG) estimator \eqref{pb:perturbed} as baseline~\cite{chang2018eusipco}. As figures of merit we adopt the root mean square error (RMSE) $\textrm{RE}_g := \|\hat\tbg - \tbg_0\|_2/\|\tbg_0\|_2$ and the support recovery accuracy $\textrm{ACC}_X:=\frac{|\text{supp}_\tau(\hat\bbX) \cap\text{supp}_\tau(\bbX_0)|}{|\text{supp}_\tau(\bbX_0)|}$, where $\textrm{supp}_\tau(\cdot)$ is the support function with threshold $\tau=0.1$. The results are shown in Fig. \ref{f:figure_main}, where (a) and (c) are $1 - \textrm{RE}_g$ for RBDoGS and \eqref{pb:perturbed}, respectively. Apparently, for the convex approach \eqref{pb:perturbed} in~\cite{chang2018eusipco,chang2024exact}, the recovery performance $1 - \textrm{RE}_g$ is decent for small $\xi$ [cf. Theorem \ref{theorem_1}], but it worsens rapidly as $\xi$ increases. Especially for a larger $\alpha$, the eigenbasis error proxy $\xi$ has greater impact on the decrease of the recovery performance of \eqref{pb:perturbed}. This is consistent with Theorem \ref{theorem_1}, since a larger $\alpha$ reduces $Q$, leading to a lower bound \eqref{eq:thm_3} for $\|\bbDelta\|_F$ (and also $\xi$). RBDoG attains markedly better RMSE. For small $\xi$, the recovery error of $\hat\tbg$ is still small -- even for larger $\alpha$. Small perturbation of $\bbV$ can be almost perfectly corrected by the RBDoGS algorithm. By comparing (a) and (c), one can conclude that Algorithm \ref{alg:smooth} significantly improves the recovery performance when $\xi$ is larger. From Fig. \ref{f:figure_main} (b) and (d), $\textrm{ACC}_X$ trends are consistent with our discussion for the RMSE.\vspace{2pt}

\noindent\textbf{Test case 2: Covariance matrix eigenvector.} To assess recovery performance in the scenario of Remark \ref{rem:covariance}, 
we generate Bernoulli-Gaussian distributed sources $\bbX_0$ with $\theta=0.15$, for different sample sizes $P$. Instead of constructing an inverse filter $\bbG_0$, we directly generate the filter $\bbH_0 = \sum_{l=0}^{L-1}h_l\bbS^{l}$ with coefficients $\bbh = \bbe_1 + \bbh'$, where $\bbe_1=[1,0,\hdots,0]^\top\in\reals^L$ for $L = 5$ and $\bbh'\sim\textrm{Normal}(\mathbf{0}_L,\bbI_L)$  (then normalized $\|\bbh'\|_2=1$). Once more, the observations are generated as $\bbY = \bbH_0\bbX_0$. 

For comparison, we also implement the BDoG approach in~\cite{chang2018eusipco} and two recent  blind deconvolution methods in~\cite{victor2024icassp} that are robust to edge perturbations, i.e., RBD-G and RBD-G-rew. 
%
Results are depicted in Fig. \ref{f:corr_network}. Apparently, RBD-G and RBD-G-rew suffer from the fact that the edgewise perturbation model is not ideal to capture the difference between $\hbC_y$ and the true GSO. Hence the RMSE in (a)-(c), i.e., $\textrm{RE}_{H}= \|\hat\bbH_0 - \bbH_0)\|_F/\|\bbH_0\|_F$, likewise $\textrm{RE}_{G}$ and $\textrm{RE}_{X}$, all attain fairly high values ($>0.1$). However, for $\textrm{ACC}_X$, both algorithms still achieve acceptable performance above 0.8. For BDoG (no eigenvector denoising), the RMSE of $\bbG,\bbH$ and $\bbX$ is moderate and around 0.1, consistent with Theorem \ref{theorem_1}. RBDoGS outperforms all of the other three baselines and consistently attains the lowest RMSE. Moreover, Fig. \ref{f:corr_network} (d) shows that it almost perfectly identifies the support of $\bbX_0$. While \eqref{pb:smooth} is non-convex, Algorithm \ref{alg:smooth} markedly reduces error with proper initialization $\bbV_p$ extracted from $\hbC_y$.

\section{Concluding Remarks}\label{sec:conclusions}

We studied robustness of blind deconvolution on networks, when required spectral graph information is imperfect. 
For the BDoG method in~\cite{chang2018eusipco,chang2024exact}, we established a stable recovery result by deriving an error bound that holds when the GSO eigenbasis perturbation is small. We also contributed a new robust formulation and associated RBDoGS algorithm to jointly recover the inverse filter and sparse latents, aided by graph eigenbasis denoising. Numerical experiments show that RBDoGS compares favorably w.r.t. state-of-the-art approaches, especially for larger perturbations. 
Ongoing and future work includes generalizations to directed graphs~\cite{marques2020digraphs}, online algorithms for streaming data, and GSP-aware supervised learning models leveraging the algorithm unrolling principle~\cite{chang2022eusipco}.


\newpage
\bibliographystyle{IEEEtran}
\bibliography{citations.bib}


\newpage


\section*{Supplementary Material}\label{S:supp_material}



\subsection*{Proof of Theorem \ref{theorem_1}}\label{App:proof_theorem}

For a given GSO $\bbS=\bbV\bbLambda\bbV^\top$, we denote a polynomial graph filter with frequency response $\tbh$ as $\ccalP(\tbh):=\bbV\diag(\tbh)\bbV^\top$. Besides, we consider the hollow matrix version of the graph filter $\ccalM(\tbh):=(\mathbf{1}_N\mathbf{1}_N^\top - \mathbf{I}_N)\circ\ccalP(\tbh)$, obtained by setting the diagonal elements of $\ccalP(\tbh)$ to zero and leaving all other entries unchanged. We are given graph signal observations $\bbY = \ccalP(\tbh_0)\bbX_0$, where $\bbX_0\in\reals^{N\times P}$ is drawn from the Bernoulli-Gaussian model with sparsity level $\theta$. Denote the support of $\bbX_0$ as $\ccalS=\textrm{supp}(\bbX_0)$, and let $\bbM^{(\ccalS)}$ be the masked matrix with entries $M_{ij}$ if $(i,j)\in\ccalS$, and $M_{ij}=0$ otherwise. 

The observed, corrupted graph is $\bbS_p=\bbV_p\bbLambda_p\bbV_p^\top$ and  we let $\bbDelta := \bbV - \bbV_p$. Recall we solve the following problem:
\begin{equation} \label{pb:perturbed_sup}
\hat{\tbg}_p  = \text{arg}\min\limits_{\tbg} \| \bbV_p\diag{(\tbg)}\bbV_p^\top\bbY\|_{1,1}
,\quad \text{s. to }\:
\mathbf{1}_N^\top\tbg = N.
\end{equation}
Here we assume the exact recovery condition for the perfectly know GSO case ($\bbDelta=\mathbf{0}_{N\times N}$) is satisfied, i.e., $\|\bbP_1^\perp \Tilde{\bbg}_0\|_2 \leq a_0$, for some $a_0\geq 0$ that does not depend on $\bbDelta$, $\bbV$, $\tbg_0$ or $N$. 
By applying the change of variables $\bbw = \tbg\circ\tbh_0$ and substituting it in $\bbY = \ccalP(\tbh_0)\bbX_0$, we have $\bbV_p\diag(\tbg)\bbV_p^\top\bbY = \ccalP(\bbw)[\bbX_0 + \bbE]$, 
where 
\begin{equation*}
{\bbE}:= (\bbV -\bbDelta)[\bbDelta^\top - \diag(\tbg_0)\bbDelta^\top\ccalP(\tbh_0)]\bbX_0.    
\end{equation*}
Then, problem \eqref{pb:perturbed_sup} can be equivalently rewritten as 
\begin{equation} \label{pb:alternate}
\hat{\bbw} = \text{arg}\min\limits_{\bbw} \| \ccalP(\bbw)[\bbX_0 + \bbE] \|_{1,1},\quad \text{s. to }\: \tbg_0^\top \bbw  = N.
\end{equation}
Note that if there is no eigenbasis perturbation, i.e., $\bbV = \bbV_p$, then $\hat\bbw = \mathbf{1}_N$ in \eqref{pb:alternate} implies $\hat{\tbg}_p = \tbg_0$ in \eqref{pb:perturbed_sup}. In the presence of a perturbation $\bbDelta\neq \mathbf{0}_{N\times N}$, $\hat\tbg_p -\tbg_0$ in \eqref{pb:perturbed_sup} can be bounded in a similar way as $\hat\bbw-\mathbf{1}_N$ in \eqref{pb:alternate}, since $\hat\tbg_p -\tbg_0=\tbg_0\circ(\hat\bbw-\mathbf{1}_N)$. Also note that a feasible $\hat\bbdelta = \hat\bbw - \mathbf{1}_N$, e.g., $\tbg_0^\top\hat\bbdelta = 0$, can be decomposed as $\hat\bbdelta = -(\bba^\top \hat{\bbb})\mathbf{1}_N + \hat{\bbb}$, where $\bba = \bbP^\perp_1\tbg_0$ and $\hat\bbb = \bbP^\perp_1\hat\bbdelta$. Then, we have $\hat{\bbw} =  [1 -(\bba^\top \hat{\bbb})]\mathbf{1}_N + \hat\bbb$. 

Our goal is to derive bounds for $\hat\bbdelta$. To this end, we have
\begin{align} \label{eq:step_1.1} 
     \| \ccalP(\hat\bbw)[\bbX_0 + \bbE] \|_{1,1} \geq {}&\|\ccalP(\hat\bbw)(\bbX_0+\bbE^{(\ccalS)})\|_{1,1}\nonumber\\
     {}& - \|\ccalP(\hat\bbw)\bbE^{(\ccalS^c)}\|_{1,1}.
\end{align}
As $\bbE^{(\ccalS)}$ has the same support as $\bbX_0$, it is sparse. From \cite[Proposition 1]{chang2024exact}, we lower bound the first summand in \eqref{eq:step_1.1}  as
\begin{equation} \label{eq:step_1.2}
        \|\ccalP(\hat\bbw)(\bbX_0+\bbE^{(\ccalS)})\|_{1,1} \geq \|\bbX+\bbE^{(\ccalS)}\|_{1,1} + PQ \|\hat\bbb\|_2,
\end{equation}
where $Q:=C_1\sqrt{\theta}\left(\sqrt{a_0^2-(1-\sigma)^2\|\bba\|_2^2}-\sigma \|\bba\|_2\right)$, with some $\sigma:=\frac{|\tbg_0^\top\bbP^\perp_1{\hat\bbb}|}{\|{\hat\bbb}\|\|\bbP^\perp_1\tbg_0\|}\in[0,1]$ and $C_1>0$. Note that we have assumed $\|\bba\|_2\leq a_0$, so $Q\geq 0$. Next, we upper bound  $\|\ccalP(\hat\bbw)\bbE^{(\ccalS^c)}\|_{1,1}$ in \eqref{eq:step_1.1} as
\begin{equation} \label{eq:step_2.1}
    \begin{aligned}
   & \|\ccalP(\hat\bbw)\bbE^{(\ccalS^c)}\|_{1,1}  = \left\|\left(1 -\bba^\top \hat{\bbb}\right)\bbE^{(\ccalS^c)} + \ccalP(\hat\bbb)\bbE^{(\ccalS^c)}\right\|_{1,1}\\
        & \leq   \|\bbE^{(\ccalS^c)}\|_{1,1}+ (a_0  \|\bbE^{(\ccalS^c)}\|_{1,1}  + \|[\bbE^{(\ccalS^c)}]^\top\bbV\odot\bbV\|_{1\rightarrow 2}) \|\hat\bbb\|_2.
    \end{aligned}
\end{equation}

Recall $\mathbf{1}_N$ should be the `ideal' perturbation-free solution of \eqref{pb:alternate}, so $N = \tbg_0^\top\mathbf{1}_N$. For optimality, we must have 
\begin{align} \label{eq:step_3}
        \| \ccalP(\hat\bbw)[\bbX_0 + \bbE] \|_{1,1} & \leq \|  \ccalP(\mathbf{1}_N)[\bbX_0 + \bbE] \|_{1,1} \nonumber\\
        & = \|\bbX+\bbE^{(\ccalS)}\|_1 + \|\bbE^{(\ccalS^c)}\|_{1,1}.
\end{align}
From \eqref{eq:step_1.1}-\eqref{eq:step_3}, we find
\begin{equation} \label{eq:step_4}
    \begin{aligned}
        \|\bbX+\bbE^{(\ccalS)}\|_{1,1} & + \|\bbE^{(\ccalS^c)}\|_{1,1}  \geq  \|\bbX+\bbE^{(\ccalS)}\|_{1,1} - \|\bbE^{(\ccalS^c)}\|_{1,1} \\
          & \hspace{-1.5cm}+ ( PQ - a_0  \|\bbE^{(\ccalS^c)}\|_{1,1}  - \|[\bbE^{(\ccalS^c)}]^\top\bbV\odot\bbV\|_{1\rightarrow 2}) \|\hat\bbb\|_2\\
    \end{aligned}
\end{equation}
and if $PQ - a_0  \|\bbE^{(\ccalS^c)}\|_{1,1}  - \|[\bbE^{(\ccalS^c)}]^\top\bbV\odot\bbV\|_{1\rightarrow 2}> 0$, then
\begin{equation}
    \|\hat\bbb\|_2  \leq \frac{2 \|\bbE^{(\ccalS^c)}\|_{1,1}}{PQ - a_0  \|\bbE^{(\ccalS^c)}\|_{1,1}  - \|[\bbE^{(\ccalS^c)}]^\top\bbV\odot\bbV\|_{1\rightarrow 2}}.
\end{equation}
Because $\hat\tbg = \tbg_0 \circ \hat\bbw$, the difference vector between $\hat\tbg$ and the `ideal' ground-truth $\tbg_0$ is $\bbd_g = \hat\tbg - \tbg_0 = \tbg_0\circ\bbw - \tbg_0 = \tbg_0\circ(\hat\bbw-\mathbf{1}_N) = \tbg_0\circ\hat\bbdelta$. Recalling that $\hat\bbdelta = - \frac{\tbg_0^\top\hat\bbb}{N} + \hat\bbb = \left(\bbI_N - \mathbf{1}_N\frac{\tbg_0^\top}{N}\right)\hat\bbb$, we can bound the $\ell_2$ norm of $\bbd_g$ as
\begin{align*}
        \|\bbd_g\|_2 & = \|\tbg_0\circ\hat\bbdelta\|_2\\
        & = \left\|\diag(\tbg_0) \left(\bbI_N - \mathbf{1}_N\frac{\tbg_0^\top}{N}\right)\hat\bbb\right\|_2\\
        &\leq \sigma_\textrm{max}\left[\diag(\tbg_0) \left(\bbI_N - \mathbf{1}_N\frac{\tbg_0^\top}{N}\right)\right]\|\hat\bbb\|_2\\
        & \leq \frac{2\sigma_\textrm{max}\left[\diag(\tbg_0) - \frac{\tbg_0\tbg_0^\top}{N}\right]\bbE^{(\ccalS^c)}\|_{1,1}}{P Q - a_0  \|\bbE^{(\ccalS^c)}\|_{1,1}  - \|[\bbE^{(\ccalS^c)}]^\top\bbV\odot\bbV\|_{1\rightarrow 2}},
\end{align*}
completing the proof.
$\hfill\blacksquare$


\end{document}